\documentstyle[amssymb,prl,aps,multicols,epsf]{revtex}

\begin{document}
\title{Quintessence and Scalar Dark Matter in the Universe}
\author{Tonatiuh Matos\thanks{%
E-mail: tmatos@fis.cinvestav.mx} and L. Arturo Ure\~na-L\'opez\thanks{%
E-mail: lurena@fis.cinvestav.mx}}
\address{Departamento de F\'{\i}sica, \\
Centro de Investigaci\'on y de Estudios Avanzados del IPN,\\
AP 14-740, 07000 M\'exico D.F., MEXICO.\\
}
\date{\today}
\maketitle

\begin{abstract}
Continuing with previous works, we present a cosmological model in which
dark matter and dark energy are modeled by scalar fields $\Phi $ and $\Psi $%
, respectively, endowed with the scalar potentials $V(\Phi )=V_{o}\left[
\cosh {(\lambda \,\sqrt{\kappa _{o}}\Phi )}-1\right] $ and $\tilde{V}(\Psi )=%
\tilde{V_{o}}\left[ \sinh {(\alpha \,\sqrt{\kappa _{o}}\Psi )}\right]
^{\beta }$. This model contains 95\% of scalar field. We obtain that the
scalar dark matter mass is $m_{\Phi }\sim 10^{-26}eV.$\ The solution
obtained allows us to recover the success of the standard CDM.  The
implications on the formation of structure are reviewed. We obtain that the
minimal cutoff radio for this model is $r_{c}\sim 1.2 \, kpc.$
\end{abstract}

\draft
\pacs{PACS numbers: 98.80.-k, 95.35.+d}

\begin{multicols}{2}  \narrowtext

For years, there has been a lot of evidence about the missing matter in the
Universe. It is known that the components of the Universe are radiation,
baryons, neutrinos, etc. but observations show that their contribution is less
than 5 \% of the total mass of the Cosmos, in agreement with Big Bang
Nucleosynthesis predictions. This suggests that there must exist a
non-baryonic type of matter in galaxies and clusters of galaxies\cite
{schram,ostriker}. Recently, the observations of Ia-type supernovae\cite
{perlmutter,riess} showed that there must exist another component that accelerates the expansion of the Universe. This new
component must have a negative equation of state $\omega <-1/3$, where $%
p=\omega \rho $\cite{stein3}. The observations point out into a flat
Universe filled with radiation, plus baryons, plus neutrinos, etc.
contributing with $\sim $ 5\%, a dark matter component with $\sim $ 25\% and
the so called dark energy contributing with $\sim $ 70\% to the total mass
of the Cosmos\cite{turner}. One of the most successful models until now is
the $\Lambda $ Cold Dark Matter ($\Lambda$CDM) model, where the dark energy
is a cosmological constant\cite{triangle,alexei}. However, some problems of this
model has not been solved yet. First of all, if a cosmological constant
exists, why is its contribution to the total matter of the same order of
magnitude as baryons and cold dark matter?. This is the cosmic coincidence problem. Also, the suggested value for the cosmological constant appears well below the values predicted by particle physics. On the other hand, the existence of a cosmological constant leads to a strong fine tuning problem over the initial conditions of the Universe.

These last facts open the possibility for the scalar fields as strong
candidates to be the missing matter of the Universe\cite{siddh,ferr,stein,luis}. A reliable model for dark energy is a fluctuating, inhomogeneous scalar field, rolling down a scalar potential, called Quintessence ($Q$)\cite{stein4}. For this case, great effort has been done to determine the appropriate scalar potential that could explain current cosmological observations\cite{ferr,stein,barr}. One example, is the pure exponential potential\cite{ferr,barr}. It has the advantages that it mimics the dominant density background and it appears naturally as a solution for a completely scalar dominated Universe\cite{urena}. But nucleosynthesis constraints require that the scalar field contribution be $\Omega _{\Phi}\leq 0.2$, which indicates that the scalar field would never dominate the Universe\cite{ferr}. However, a special group of scalar potentials has been
proposed in order to avoid the fine tuning and coincidence problem: the
tracker solutions\cite{stein}, where the cosmology at late times is
extremely insensitive to initial conditions. A typical potential is the pure
inverse power-law one, $V(\Phi )\sim \Phi ^{-\alpha }$ ($\alpha >0$)\cite
{stein,peebles2}. Although it reduces the fine tuning and the cosmic
coincidence problem, the predicted value for the current equation of state
for the quintessence is not in good agreement with supernovae results\cite
{stein}. The same problem arises with the inverse power-law-like potentials.
Another example are the potentials proposed in \cite{varun}. They avoid
efficiently the troubles stated above, but it is not possible to determine
unambiguously their free parameters.

In this letter we use a potential $\cosh $, in order to mimic a standard
cold dark matter with a quintessential dark energy. Then we will investigate
the scalar field fluctuations and the implications in structure formation in
directions suggested by some authors. We obtain that the scalar field is a
ultra-light particle which behaves just as cold dark matter. Using previous
works \cite{siddh,ann,sph}, it is then possible that a scalar field
fluctuation could explain the formation of the galaxy halos.

In a recent paper\cite{luis}, we showed that the potential

\begin{eqnarray}
\tilde{V}(\Psi ) &=&\tilde{V_{o}}\left[ \sinh {(\alpha \,\sqrt{\kappa _{o}}%
\Psi )}\right] ^{\beta }  \label{sinh} \\
&=&\left\{ 
\begin{array}{cc}
\tilde{V_{o}}\left( \alpha \,\sqrt{\kappa _{o}}\Psi \right) ^{\beta } & 
|\alpha \,\sqrt{\kappa _{o}}\Psi |\ll 1 \\ 
\left( \tilde{V_{o}}/2^{\beta }\right) \exp {\left( \alpha \beta \,\sqrt{%
\kappa _{o}}\Psi \right) } & |\alpha \,\sqrt{\kappa _{o}}\Psi |\gg 1
\end{array}
\right. ;  \nonumber
\end{eqnarray}

\noindent is a good candidate for the dark energy. Its asymptotic behavior
at early (late) times is the attractive inverse power-law (exponential) one.
Its parameters are given by

\begin{eqnarray}
\alpha &=&\frac{-3\omega _{\Psi }}{2\sqrt{3(1+\omega _{\Psi })}},  \nonumber
\\
\beta &=&\frac{2\,(1+\omega _{\Psi })}{\omega _{\Psi }},  \label{tracpsi} \\
\rho _{o\Psi } &=&\left( \frac{2\,\tilde{V_{o}}}{1-\omega _{\Psi }}\,\rho
_{oCDM}^{-\beta /2}\right) ^{\frac{1}{1-\,\beta /2}};  \nonumber
\end{eqnarray}

\noindent where $\rho_{oCDM}$ and $\rho_{o\Psi}$ are the current energy densities of cold dark matter and dark energy, respectively, and $\omega_\Psi$ is the current equation of state for the dark energy. It eliminates the fine tuning problem and dominates only at late times, driving the Universe to a power-law inflationary stage (for which the scale factor $a \sim t^p$, with $p>1$). Thus, again, we will take it as our model for the dark energy.

At the same time, there exists strong evidence for the scalar fields to be
the dark matter at galactic level. If the dark matter component is the
scalar field, then it was demonstrated in \cite{siddh} that a scalar field
fluctuation could behave in exactly the same way as the halo of a galaxy.
The halos of galaxies (the scalar field fluctuations) could be axial
symmetric\cite{ann} or spherically symmetric\cite{sph}, in both cases, the
geodesics of exact solutions of the Einstein equations with an exponential
potential fit the rotation curves of galaxies quite well. Besides, the $%
\Lambda$CDM model over predicts subgalactic structure and singular cores of
the halos of galaxies\cite{galaxs}. In order to solve these problems, some
authors have proposed power-law and power-law-like scalar potentials\cite
{varun,peebles,jeremy,stein2} to be the dark matter in the Universe, and it
is worth to mention that some of them could be tracker solutions by
themselves\cite{stein2}. All attention has been put on the quadratic
potential $\Phi ^{2}$, because of the well known fact that it behaves as
pressureless matter due to its oscillations\cite{turner1}, implying that $%
\omega _{\Phi}\simeq 0$, for $<p_{\Phi }>=\omega _{\Phi }<\rho _{\Phi }>$. A
reliable model for the dark matter can then be the potential ( \cite
{alexei,varun} and references therein)

\begin{eqnarray}
V(\Phi ) &=&V_{o}\left[ \cosh {(\lambda \,\sqrt{\kappa _{o}}\Phi )}-1\right]
\label{cosh} \\
&=&\left\{ 
\begin{array}{cc}
\left( V_{o}/2\right) \left( \lambda \sqrt{\kappa _{o}}\Phi \right) ^{2} & 
|\lambda \,\sqrt{\kappa _{o}}\Phi |\ll 1 \\ 
\left( V_{o}/2\right) \exp {\left( \lambda \,\sqrt{\kappa _{o}}\Phi \right) }
& |\lambda \,\sqrt{\kappa _{o}}\Phi |\gg 1
\end{array}
\right. ;  \nonumber
\end{eqnarray}

\noindent because this potential joins together the attractive properties of
an exponential potential and the already mentioned quadratic potential, as
it can be seen from its asymptotic behavior.

We consider a flat, homogenous and isotropic Universe. Thus we use the flat
Friedmann-Robertson-Walker (FRW) metric

\begin{equation}
ds^2=-dt^2+a^2(t)\left[ dr^2 + r^2 \left(d\theta^2 + \sin^2{(\phi)}d\phi^2
\right) \right] .
\end{equation}

The components of the Universe are baryons, radiation, three species of
light neutrinos, etc., and two minimally coupled and homogenous scalar
fields $\Phi $ and $\Psi $, which represent the dark matter and the dark
energy, respectively. Thus, the evolution equations for this Universe are

\begin{eqnarray}
H^{2} \equiv \left( \frac{\dot{a}}{a} \right)^2 &=&\frac{\kappa _{o}}{3}%
\left( \rho +\rho _{\Phi }+\rho _{\Psi }\right) ,  \nonumber \\
{\dot{\rho}}+3H\left( \rho +p\right) &=&0,  \nonumber \\
\ddot{\Phi}+3\dot{H}\dot{\Phi}+\frac{dV(\Phi )}{d\Phi } &=&0,  \label{evol}
\\
\ddot{\Psi}+3\dot{H}\dot{\Psi}+\frac{d\tilde{V}(\Psi )}{d\Psi } &=&0, 
\nonumber
\end{eqnarray}

\noindent being $\kappa _{o} \equiv 8\pi G$, $\rho $ the energy density of
radiation, plus baryons, plus neutrinos, etc., $\rho _{\Phi }=\frac{1}{2}%
\dot{\Phi}^{2}+V(\Phi )$ and $\rho_{\Psi }=\frac{1}{2}\dot{\Psi}^{2}+\tilde{V%
}(\Psi )$.

We start the evolution of the Universe in the radiation dominated era (RD),
with large (small) and negative (positive) values for the scalar field $\Phi 
$ ($\Psi $). Taking the initial condition $\rho _{i\Psi }<\rho _{i\gamma }$,
the energy density $\rho _{\Psi }$ is subdominant and behaves as a
cosmological constant. The tracker solution~(\ref{tracpsi}) will be reached
only until the matter dominated era (MD), that is, until the background
equation of state becomes $\omega _{b}=0$. Since then, it will evolve with a
constant equation of state $\omega _{\Psi }$ and will dominate the current
evolution of the Universe as an effective exponential potential. The
Universe would then be in a power-law inflationary stage. More details can
be found in\cite{luis}. Now, we will focus our attention in potential~(\ref
{cosh}).

During RD, the scalar field energy density $\rho _{\Phi }$ tracks the
radiation energy density. More, the ratio of $\rho_\Phi$ to the total energy
density is constant and equals to (see\cite{ferr,chimen} and references
therein):

\begin{equation}
\frac{\rho _{\Phi }}{\rho _{\gamma }+\rho _{\Phi }}=\frac{4}{\lambda ^{2}}.
\label{rad}
\end{equation}

\noindent with $\rho _{\gamma }$ the contribution due to radiation. In order
to recover the success of the CDM model, we will make the scalar energy
follow the standard cold dark matter at the epoch of its oscillations. Thus,
we investigate the behavior of the scalar field $\Phi $ near to the
transition point $|\sqrt{\kappa _{o}}\lambda \Phi |=1$ (see eqs.~(\ref{cosh}%
)), i.e., the point when the scalar field is leaving the radiation
solution (exponential-like potential) and entering the dust solution
(quadratic-like potential). Taking $a_{\ast }$ as the value for the scale
factor when this transition occurred (the scale factor has been normalized
to $a=1$ today), we find that it can be approximately given by

\begin{equation}
a_{\ast }\approx \frac{4}{\lambda ^{2}-4}\left( \frac{\Omega _{o\gamma }}{%
\Omega _{oCDM}}\right) .  \label{ast}
\end{equation}

\begin{figure}[h]
\centerline{ \epsfysize=5cm \epsfbox{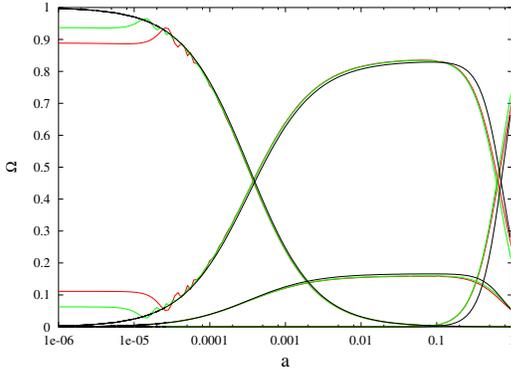}}
\caption{Evolution of the dimensionless density parameters {\it vs} the scale factor $a$ with $\Omega_{oM} = 0.30$: $\Lambda$CDM (black) and $\Psi \Phi DM$ for two values of $\protect\lambda=6$ (red), $\protect\lambda=8$ (green). The equation of state for the dark energy is $\protect\omega_\Psi = -0.8$.}
\label{fig:Omegac}
\end{figure}

From this, it can be shown that for potential (\ref{cosh}) it follows

\begin{equation}
\frac{\kappa_o V_o}{\left( \lambda^2 -4\right)^3} \simeq \frac{1.7}{3} \left[
\left(\frac{\Omega_{oCDM}}{\Omega_{o\gamma}} \right)^3 \Omega_{oCDM} \right]
H^2_o.  \label{ratio}
\end{equation}

\noindent being $\Omega _{oCDM}$ and $\Omega _{o\gamma }$ the current
measured values for the densities of dark matter and radiation,
respectively, and $H_{o}$ the current Hubble parameter. The restriction from
nucleosynthesis for the early exponential behavior of the potential requires

\begin{equation}
\frac{\rho_\Phi}{\rho_\gamma} = \frac{4}{\lambda^2 -4} < 0.2  \label{lambda}
\end{equation}

\noindent at the radiation dominated era\cite{ferr}. Then we have that $%
\lambda >2\sqrt{6}$. A numerical solution for the density parameters $\Omega
_{X}=(\kappa _{o}\rho _{X})/(3H^{2})$ is shown in fig.~(\ref{fig:Omegac}).
The time when oscillations start is well given by eq.~(\ref{ast}), and with
the values from eq.~(\ref{ratio}) the solution mimics quite well the
standard CDM model until today (see for example\cite{luis}). Note that
the change of $\rho _{\Phi }$ to a dust solution occurred before the
radiation-matter equality for the values given by eq.~(\ref{ratio}). This
allows the scalar field $\Phi $ to dominate the evolution of the Universe
later, and to provoke a MD era\cite{picci}.

\begin{figure}[h]
\centerline{ \epsfysize=5cm \epsfbox{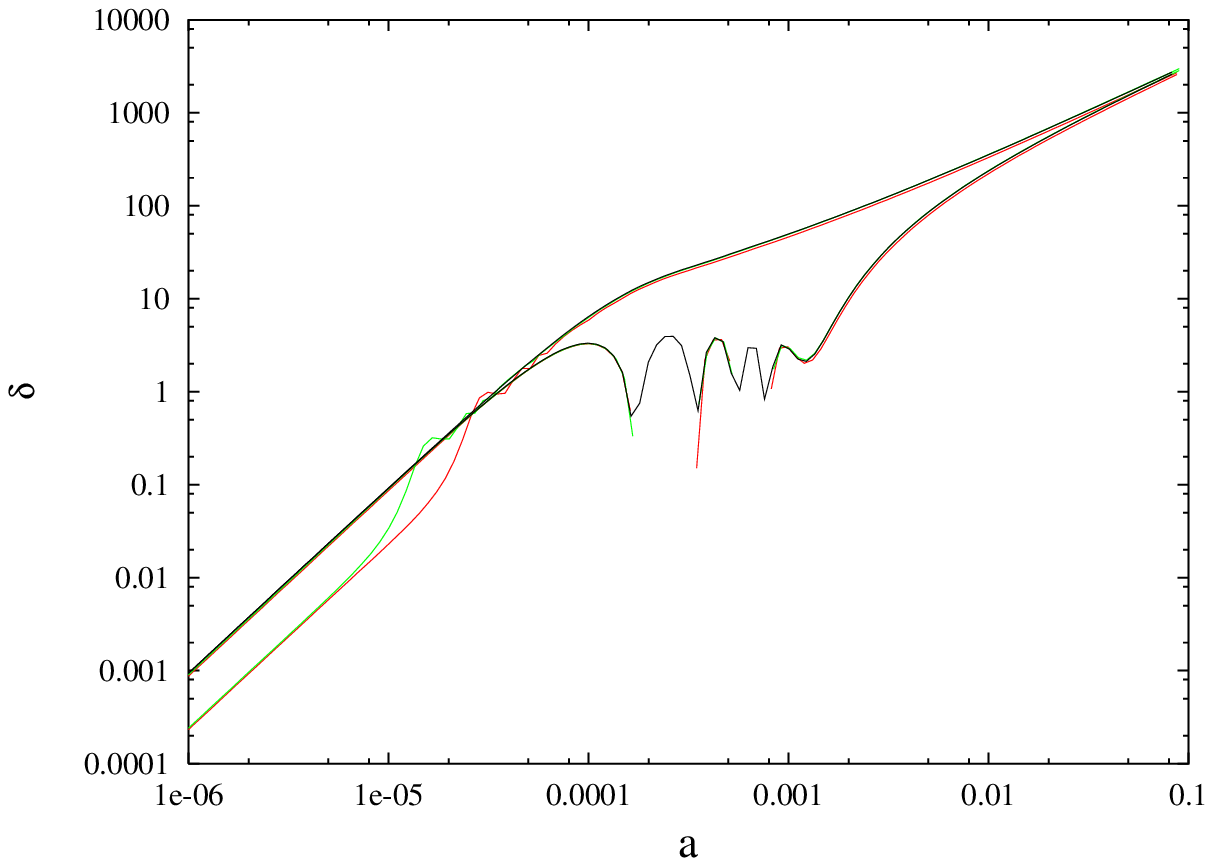}}
\centerline{ \epsfysize=5cm \epsfbox{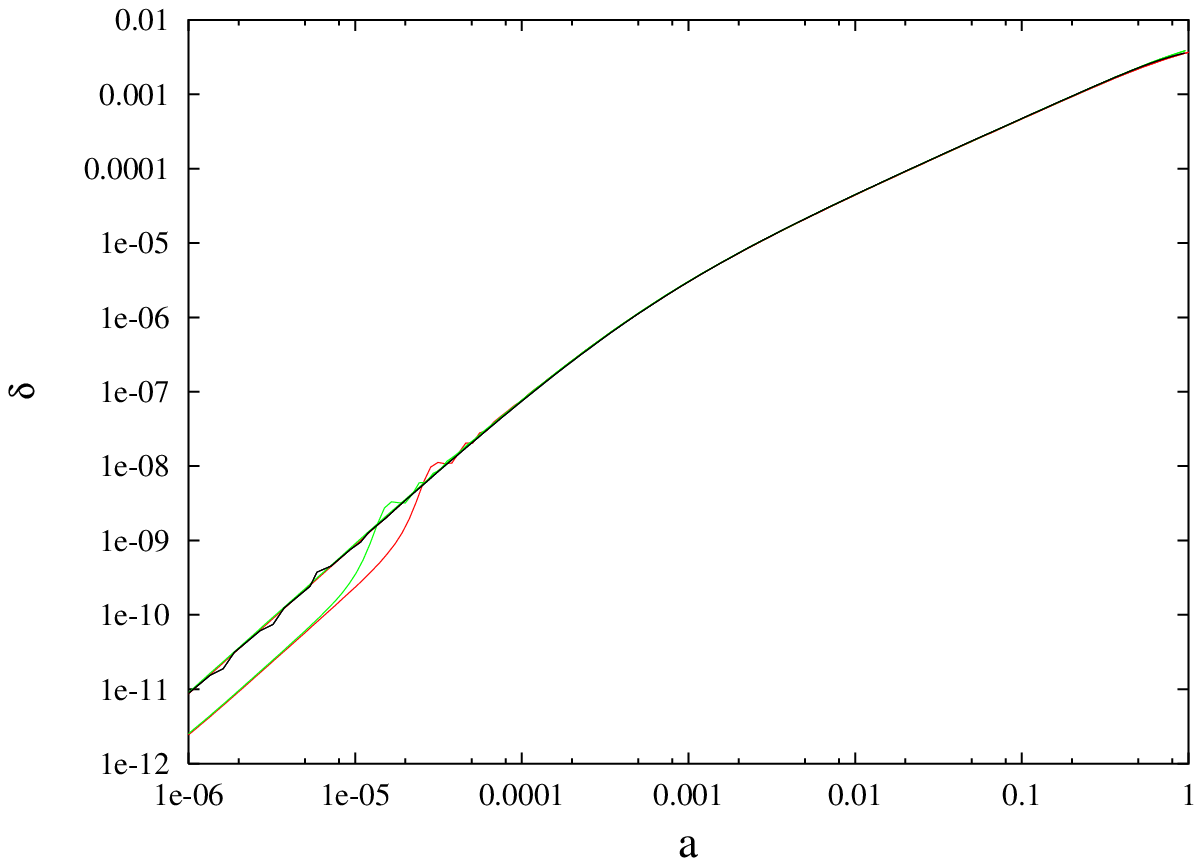}}
\caption{Evolution of the density contrasts $\delta_b$ (baryons), $\delta_{CDM}$ (standard cold dark matter) and $\delta_\Phi$ (scalar dark matter) {\it vs} the scale factor $a$ for the models given in fig.~(\ref{fig:Omegac}). The modes shown are $k=0.1 \, Mpc^{-1}$ (top) and $k=1.0 \times 10^{-5} \, Mpc^{-1}$ (bottom).}
\label{fig:dphi}
\end{figure}

Now we will investigate the fluctuations in the scalar dark matter
component. Using an amended version of CMBFAST\cite{seljak} and taking
adiabatic initial conditions\cite{ferr}, we observe that the scalar
fluctuations of $\Phi $ make the scalar density contrast $\delta _{\Phi
}=\left( \delta \rho _{\Phi }/\rho _{\Phi }\right) $ follow the standard
dark matter density contrast (see fig.~(\ref{fig:dphi}))\cite{ma}. We have
then a kind of tracker solution for the fluctuations of the scalar dark
matter, too. This last fact makes the potential~(\ref{cosh}) a reliable dark
matter one.

The mass for the scalar field $\Phi$ is $m^2_\Phi = V^{\prime \prime} (0)=
\kappa_o V_o \lambda^2$. Observe that from eq.~(\ref{lambda}) we have a
minimal value for the mass of the field. Using eq.~(\ref{ratio}) we get

\begin{equation}
m^2_{\Phi,min} \simeq 1.08 \times 10^5 \left[ \left(\frac{\Omega_{oCDM}}{%
\Omega_{o\gamma}} \right)^3 \Omega_{oCDM} \right] H^2_o.  \label{minmass}
\end{equation}

\noindent implying that $m_{\Phi }>3\times 10^{-26}\,eV$; thus, we are
dealing with an ultra-light particle as dark matter. Since the Compton length
is related to the mass by $\lambda _{C}=m_{\Phi }^{-1}$, there will be a
maximum value for $\lambda _{C}$ given by

\begin{equation}
\lambda_{C,max} \simeq 3.0 \times 10^{-3} \left[ \left(\frac{\Omega_{oCDM}}{%
\Omega_{o\gamma}} \right)^3 \Omega_{oCDM} \right]^{-1/2} H^{-1}_o,
\end{equation}

\noindent and then $\lambda _{C}<200\,pc$.

From eq. (\ref{ratio}), we can see that there is a degeneracy because of the
infinite pairs ($V_{o},\lambda $) that are available for the same values of $%
\Omega _{oCDM}$ and $\Omega _{o\gamma }$, and that we recover the standard
dark matter model if $\lambda \rightarrow \infty $. In fact, observe that we
have a one-parameter theory where we can chose $V_{o}$, $\lambda $ or $%
m_{\Phi }$ as a free parameter. Then, we need another observational
constraint to fix completely the parameters of the potential. In\cite{varun}%
, it was suggested that the Compton length could be a cutoff for structure
formation, but its value is not big enough to be useful. In\cite{jeremy}, it
is studied a similar model but here the scalar particles behave like a
relativistic gas before the time of radiation-matter equality, the gas being
non-relativistic at the current epoch. This last fact ensures that the
minimal scale for dark matter halos is of order of $kpc$. The potential used
in \cite{jeremy} is

\begin{equation}
V(\Phi )=\frac{m_{\Phi }^{2}}{2}\Phi ^{2}+\kappa \Phi ^{4}  \label{jeremyp}
\end{equation}

\noindent being $\kappa $ the dimensionless free parameter of the model. For
the potential (\ref{cosh}), $\kappa$ is no longer a free parameter, but $%
\kappa =\kappa _{o}\lambda ^{2}m_{\Phi }^{2}/4!$. Then, in our case the
minimal radius for compact equilibrium now reads \cite{jeremy}

\begin{equation}
r_c=(3/2)\left( \kappa _{o}V_{o}\right) ^{-1/2}.
\end{equation}
Taking the minimal value for $\lambda$ allowed by nucleosynthesis and using
eq.~(\ref{ratio}), the available values for $r_c$ are

\begin{equation}
r_c \leq 2 \times 10^{-2} \left[ \left(\frac{\Omega_{oCDM}}{%
\Omega_{o\gamma}} \right)^3 \Omega_{oCDM} \right]^{-1/2} H^{-1}_o,
\end{equation}

\noindent thus $r_c \leq 6\lambda _{C}\simeq 1\,.2kpc$. This value can
be useful in order to explain the suppression of galactic substructure and
could give us the new constraint we need to fix all the parameters of the
model.

Summarizing, a model for the Universe where 95\% of the energy density is of
scalar nature can be possible. This would have strong consequences in
structure formation, like the suppression of subgalactic objects due to the
dark matter composed of a ultra-light particle.

\acknowledgements{We would like to thank F. Siddhartha Guzm\'an and Dario
Nu\~nez for helpful discussions. This work was partly supported by CONACyT,
M\'{e}xico 119259 (L.A.U.)}

\end{multicols}

\end{document}